\documentclass[letterpaper,12pt]{article}

\usepackage{fancyhdr}
\usepackage{amsmath}
\usepackage{amsbsy}
\usepackage{amssymb}
\usepackage{amsthm}
\usepackage{amscd}
\usepackage{amsfonts}
\usepackage{supertabular}
\usepackage{graphics}
\usepackage{graphicx}
\usepackage{verbatim}
\usepackage{subfigure}
\usepackage{epsfig}
\usepackage{xspace}
\usepackage{euscript}
\usepackage{alltt}
\usepackage{boxedminipage}
\usepackage{float}
\usepackage{times}
\usepackage[colorlinks]{hyperref}
\usepackage[round,authoryear]{natbib}
\usepackage{setspace}
\usepackage{amsmath, amssymb}
\usepackage{algorithm}
\usepackage{booktabs}
\usepackage{algorithmic}
\usepackage{lscape}
\usepackage{wrapfig}
\usepackage{subfigure}%
\usepackage[hang,small,bf]{caption}
\usepackage{color}
\usepackage{authblk}
\usepackage[pagewise,displaymath]{lineno}

\def\Put(#1,#2)#3{\leavevmode\makebox(0,0){\put(#1,#2){#3}}}

\begin{document}


\title{Toward predicting tensile strength of pharmaceutical tablets by ultrasound measurement in continuous manufacturing}

\author[a]{Sonia M. Razavi}
\author[a]{Gerardo Callegari}
\author[a]{German Drazer}
\author[a]{Alberto M. Cuiti\~{n}o \thanks{Corresponding author at: Department of Mechanical and Aerospace Engineering, Rutgers University, Piscataway, NJ 08854, USA. Tel.:~+1~848 445 2248~. Fax: +1~732 445 3124 \\ \indent E-mail address: cuitino@jove.rutgers.edu}}
\affil[a]{\small Department of Mechanical and Aerospace Engineering, Rutgers, The State University of New Jersey, Piscataway, NJ 08854, USA}

\maketitle

\begin{abstract}

An ultrasound measurement system was employed as a non-destructive method to evaluate its reliability in predicting the tensile strength of tablets and investigate the benefits of incorporating it in a continuous line, manufacturing solid dosage forms. Tablets containing lactose, acetaminophen, and magnesium stearate were manufactured continuously and in batches. The effect of two processing parameters, compaction force and level of shear strain were examined. Young's modulus and tensile strength of tablets were obtained by ultrasound and diametrical mechanical testing, respectively. It was found that as the blend was exposed to increasing levels of shear strain, the speed of sound in the tablets decreased and the tablets became both softer and mechanically weaker. Moreover, the results indicate that two separate tablet material properties (e.g., relative density and Young's modulus) are necessary in order to predict tensile strength. A strategy for hardness prediction is proposed that uses the existing models for Young's modulus and tensile strength of porous materials. Ultrasound testing was found to be very sensitive in differentiating tablets with similar formulation but produced under different processing conditions (e.g., different level of shear strain), thus, providing a fast, and non-destructive method for hardness prediction that could be incorporated to a continuous manufacturing process.
\end{abstract}


\section{Introduction}

\hspace{5 mm} There are many advantages to Continuous Manufacturing (CM), including: (i) integrated processing with fewer steps, which results in minimal manual handling and increased safety; (ii) smaller facilities (i.e., reduced cost); and (iii) on-line monitoring and control for enhanced product quality assurance in real-time \citep{chatterjee2012fda}. Many industries, such as petrochemical and food technologies, have shifted to CM. However, in the pharmaceutical industry, products are still manufactured mainly in batches. The major barriers to CM are traceability and the lack of rapid techniques for quality assurance and control. In an attempt to explore and address the manufacturing issues in the pharmaceutical industry, the U.S. Food and Drug Administration (FDA) has released the Process Analytical Technology (PAT) initiative. Designing and developing rapid techniques and ultimately improving the quality of pharmaceutical products are the goals of the PAT initiative \citep{food2004guidance}. 

Among pharmaceutical products, tablets are the most common dosage form due to their high production rates, acceptable shelf life, dosage accuracy, and contolled drug release. The mechanical strength of tablets is an important quality attribute that is consistently tested to ensure that tablets can withstand post-compaction operations, such as coating, handling, and storage. The dissolution profile of a drug tablet is also influenced by its mechanical properties \citep{saravanan2002effect}. Lubrication, among other factors, may significantly affect the mechanical properties of tablets\citep{johansson1984granular}. Lubricants are an essential ingredient in tablet formulations to prevent powders from sticking to the tooling and improve powder flow properties during the compaction process \citep{moody1981tablet}. Concentration of lubricant and exposure to shear are two important variables in the lubrication process. A significant reduction in tablet hardness due to overlubrication with magnesium stearate (MgSt), for example, has been previously shown \citep{bolhuis1975film,de1978bonding,bossert1980effect, kikuta1994effect}, and is caused by the formation of MgSt film on powder particles, which weakens the interparticle bonding \citep{bolhuis1975film,de1978bonding}.

The mechanical strength of tablets is typically measured by traditional destructive tests, such as three-point bending, four-point bending, diametrical compression, and axial tensile strength tests \citep{stanley2001mechanical,podczeck2012methods}. These destructive tests not only damage the tablet structure and cause loss of product, they also provide limited information about the mechanical state of a tablet. Moreover, the time spent to test the tablets destructively is in the order of minutes, which is not suitable for an on-line monitoring process. Ultrasound (US) testing has been recently introduced as a fast, non-invasive technique to measure tablet strength. This requires measuring the time of flight (TOF) of a low intensity mechanical wave propagating in the tablet. The Young's modulus (E) can then be calculated by determining the longitudinal speed of sound (SOS) of this transmitted US signal from the TOF measured in a tablet of known thickness \citep{akseli2009non}. 
\cite{hakulinen2008ultrasound} have observed that SOS decreases as the porosity of the tablet increases. In addition, the SOS in a tablet was found to increase with its tensile strength \citep{akseli2011quantitative,simonaho2011ultrasound}. The ease of implementation, fast computing time, and low cost of this method make it possible to be placed on-line for real-time mechanical characterization of tablets. \cite{leskinen2010line} have introduced an in-die US measurement system by incorporating US transducers inside the upper and lower punches. US attenuation was found to be a good approach to detect defective tablets. In a later study, they measured the SOS in binary mixtures (i.e., mixtures of an active ingredient with an excipient) during tableting using the same system. They found that SOS is sensitive to the mixing time of magnesium stearate and the dwell time of the compaction cycle \citep{leskinen2013real}. The in-die real-time tablet monitoring system has also been used by \cite{stephens2013ultrasonic} to evaluate the tablet mechanical integrity and the presence of defects, and its applicability as a control system was validated. Although the in-die measurement provides valuable information, the mechanical strength of a tablet is different if measured out-of-die. After compaction and in-die unloading the tablet experiences ejection forces, as well as radial and axial elastic relaxation, which might siginificantly affect the mechanical integrity of the tablet \citep{train1956investigation,long1960radial,maarschalk1998influence}.

In this study, we focused on evaluating the mechanical integrity of tablets after compaction via US testing. Cylindrical tablets were prepared either continuously or in batch. The formulation was kept constant ($90\%$ lactose monohydrate, $9\%$ acetaminophen (APAP), and $1\%$ MgSt), while the compaction force and level of shear strain varied. US testing was used to evaluate the strength of tablets by measuring the TOF. The tensile strength of the same tablets was then determined using a mechanical hardness tester. It was observed that, as the blend was exposed to an increasing level of shear strain, the speed of sound decreased and the tablets became both softer and mechanically weaker. It is also noticed that in order to predict the hardness of a tablet, two properties should be taken into account: Young's modulus and relative density. A strategy for hardness prediction is proposed that uses the existing theoretical/semi-empirical models for Young's modulus and tensile strength of porous materials. Overall, US testing is found to be a reliable technique to predict the variation of tablet strength with processing conditions.

\section{Materials and methods}
\label{sec:M&M}

\subsection{Materials}

\hspace{5 mm}Lactose (monohydrate N.F., crystalline, $310$, Regular, Foremost Farms USA, Rothschild, Wisconsin, USA), acetaminophen (semi-fine, USP/paracetamol PhEur, Mallinckrodt, Raleigh, North Carolina, USA), and magnesium stearate N.F. (non-Bovine, Tyco
Healthcare / Mallinckrodt, St. Louis, Missouri, USA) were used as purchased. A formulation containing $90\%$ lactose, $9\%$ acetaminophen (APAP), and $1\%$ magnesium stearate (MgSt) was prepared on a weight basis and kept constant in both batch and continuous production.

\begin{table}[htb]
    \centering
\caption{Blend constituents, nominal mean particle size, and true density.}
    \scriptsize\setlength{\tabcolsep}{8pt}{
    \begin{tabular}{c   ccc  ccc} 
      \hline
         Material & Mean particle size ($\mu$$m$) & True density ($g/cm\textsuperscript{3}$) 
    \\
\hline
    Lactose & 180 & 1.56
    \\
   Acetaminophen (APAP) & 45 & 1.30
    \\
   Magnesium Stearate& 10 & 1.04
    \\
\hline
    \end{tabular}
    }
    \label{Table:particel size}
\end{table}

\hspace{-2mm}The true density was measured with five parallel measurements with a pycnometer (AccuPyc $1340$, Micromeritics) using helium as the measuring gas. The nominal particle sizes and true densities of the materials used are listed in Table \ref{Table:particel size}.

\subsection{Continuous Manufacturing}

\hspace{5 mm}The pilot plant employed for continuous manufacturing is situated at the Engineering Research Center for Structured Organic Particulate Systems (ERC-SOPS), Rutgers University. A detailed description of this plant can be found in \cite{singh2014systematic}. There are four main unit operations when tablets are produced by direct powder compression: feeding, delumping, blending, and compacting. We will present a brief overview of each of these operations next.

\textit{Feeding and delumping}: First, from gravimetric feeders (K-Tron KT20) APAP and lactose were separately fed into a mill (Quadro Comil 197-S). The MgSt was added after the mill to prevent overlubrication of the formulation. 

\textit{Blending}: The blend was then sent to a continuous blender (Glatt GCG-70) with a speed set at $200$rpm. Chemical composition of the powder was monitored using a Bruker Matrix near-infrared (NIR) spectrometer. 

\textit{Compaction}: The desired formulation was sent through a hopper into the fill-o-matic with a speed set at $25$rpm. Finally, the blend was compressed using a 36-station Kikusui Libra-$2$ double layer tablet press with a $10$mm tooling at a compaction speed of $20$rpm. 

The overall flow rate and the tablet weight were set to $20$kg/h and $350$mg, respectively. The tablets made in the continuous line will be referred to as \textit{continuous} tablets. The compaction force ($\text{F}_c$) was varied during the run by changing the distance between punches. Since the individual $\text{F}_c$ values are not recorded for each tablet, we categorize the tablets based on their nominal compaction force values ($\text{F}_n$). Tablets were collected after all the processing parameters reached steady state. $6$ different $\text{F}_n$ settings were selected ranging from $8$ to $28$kN. $12$ tablets for each of $8$, $20$, and $28$kN and $18$ tablets for each of $12$, $16$, and $24$kN $\text{F}_n$ conditions were analyzed.

\subsection{Batch Production}

\subsubsection{Blend Preparation}

\hspace{5 mm}Two different powder mixing equipments were used in the batch production of tablets: a V-blender  (Patterson-Kelley Co., East Stroudsburg, PA) and a laboratory scale resonant acoustic mixer (labRAM) (Resodyn Acoustic Mixers, Butte, Montana, USA). 

In the V-blender mixer, a $15$-minute pre-blending step was applied at a rotation rate of $15$ rpm to reduce the stickiness of APAP and improve its flowability. MgSt was added and mixed with the blend for $2$ additional minutes. The blended powder was then unloaded from the V-blender and subjected to a controlled shear environment in a modified Couette cell at a shear rate of $80$ rpm. For additional information about this instrument, the reader is referred to \cite{mehrotra2007influence} and \cite{pingali2011mixing}. Three different shear strain environments were selected in this study: $0$, $160$ and $640$ revolutions, corresponding to $0$, $2$, and $8$ minutes in the shear device. As we increase the number of revolutions in the shear cell, the degree of MgSt coverage on the particles increases. The tablets made using this mixing method will be named according to the strain
level experienced by the blend in the shear cell i.e., \textit{$0$rev}, \textit{$160$rev}, and \textit{$640$rev} tablets.

The low frequency and high intensity acoustic field in labRAM facilitate the movement of the loose powder mass to induce mixing. Fill level, blending time, and acceleration (mixing intensity) are the key parameters that affect the labRAM performance \citep{osorio2015evaluation}. In the labRAM mixer, $80\%$ fill level was selected. Lactose and APAP were first blended for $2$ minutes at an acceleration of $40$g, followed by the lubrication stage, where MgSt was added and the blend was mixed for an additional $1$ minute at an acceleration of $60$g.  These tablets will be referred to as \textit{labRAM} tablets.

\subsubsection{Tablet Compaction}

\hspace{5 mm}The blends were compacted using a Presster tablet press simulator (The Metropolitan Computing Corporation of East Hanover, NJ) equipped with a $10$mm flat-face, B-type tooling. A Kikusui Libra-$2$ tablet press was emulated at a press speed of $20$rpm. A dwell time of $22.2$ms, corresponding to a production speed of $43~100$ tablets per hour, was used. No pre-compression force was applied. These parameters in the Presster (e.g., tooling type and dwell time) were selected based on those employed in the tablet press of the continuous line. A total of $36$ tablets for $0$rev, $160$rev, and $640$rev conditions and $25$ labRAM tablets were analyzed.

\begin{table}[htb]
    \centering
\caption{Compaction force ($\text{F}_c$), Weight (W), thickness ($t$), relative density ($\bar{\rho}$), speed of sound (SOS), tensile strength ($\sigma_t$), and Young's modulus (E) of the labRAM tablets.}
  \scriptsize\setlength{\tabcolsep}{8pt}{
    \begin{tabular}{c   ccccccc} 
      \hline
         $\text{F}_c$ (kN) & W (mg) & $t$ (mm) & $\bar{\rho}$ (\%) & SOS (m/s) & $\sigma_t$ (MPa) & E (GPa)
    \\
\hline
     5.2 & 359.2 & 3.67 & 81.7 & 614  & 0.24 & 0.47
    \\
     6.1 & 352.8 & 3.55 & 83 & 677 & 0.28 & 0.58 
    \\
     6.6 & 344.6 & 3.43 & 83.9 & 681 & 0.33 & 0.59
    \\
     10.6 & 349.5 & 3.34 & 87.4 & 823   & 0.64 & 0.9
    \\
   8.3 & 354.4 & 3.46 & 85.5 & 783  & 0.43 & 0.8
    \\
  8.6 & 349.6& 3.4 & 85.9 & 787 & 0.48 & 0.81
    \\
  6.6 & 348.5& 3.46 & 84.1 & 721 & 0.32 & 0.67
    \\
     10.3 & 343.3 & 3.28 & 87.4 & 808  & 0.61 & 0.87
    \\
     13.5 & 355.3 & 3.34 & 88.8 & 831  & 0.88 & 0.93
    \\
     12.1 & 353.2 & 3.33 & 88.6 & 816  & 0.81 & 0.9
    \\
     11 & 345& 3.28 & 87.8 & 812  & 0.72 & 0.88
    \\
   15.6 & 350.7 & 3.24 & 90.4 & 920 & 1.06 & 1.17
    \\
   15.6 & 345.4 & 3.21 & 89.8 & 907  & 1.13 & 1.13
    \\
  21 & 350.4 & 3.17 & 92.3 & 955  & 1.48 & 1.28
    \\
  22.9 & 348.9 & 3.15 & 92.5 & 972 & 1.66 & 1.33
    \\
  29.5 & 354.3 & 3.14 & 94.2 & 1006  & 2.09 & 1.45
    \\
    23.6 & 352.9 & 3.18 & 92.7 & 1013 & 1.83 & 1.45
    \\
     23.1 & 357.7 & 3.22 & 92.7 & 1006  & 1.74 & 1.43
    \\
     24.8 & 353.7& 3.17 & 93.2 & 991  & 1.99 & 1.39
    \\
     19.5 & 351.1 & 3.2 & 91.6 & 958 & 1.44 & 1.28
    \\
   17.5 & 353.7 & 3.26 & 90.6 & 931  & 1.26  & 1.2
    \\
  15.9 & 359.7 & 3.35 & 89.6 & 936 & 1.06 & 1.2
    \\
  9.5 & 336.5& 3.26 & 86.2 & 799 & 0.55 & 0.84
    \\
   23.2 & 340.7 & 3.06 & 93 & 974  & 1.77 & 1.35
    \\
  25.3 & 346.2 & 3.11 & 92.9 & 972  & 1.85 & 1.34
    \\
\hline
    \end{tabular}
    }
    \label{Table:labram}
\end{table}

\subsection{Tablet Characterization}

\hspace{5 mm}All the compacted tablets were stored at ambient room temperature and inside a sealed, clear plastic bag and kept for at least one week to allow for elastic relaxation prior to any characterization. 

\begin{table}[htb]
    \centering
\caption{Compaction force ($\text{F}_c$), Weight (W), thickness ($t$), relative density ($\bar{\rho}$), speed of sound (SOS), tensile strength ($\sigma_t$), and Young's modulus (E) of the batch tablets mixed in the V-blender and then experienced different shear strain environments.}
   \scriptsize\setlength{\tabcolsep}{8pt}{
    \begin{tabular}{c c  ccccccc} 
      \hline
 & $\text{F}_c$ (kN) & W (mg) & $t$ (mm) & $\bar{\rho}$ (\%) & SOS (m/s) & $\sigma_t$ (MPa) & E (GPa)
    \\
\hline
 $0$rev &    8.2 & 318.4 & 3.14 & 84.7 & 969  & 0.74 & 1.21
    \\
     &10.7 & 343.1 & 3.32 & 86.3 & 1012 & 1.01 & 1.35 
    \\
   & 14.3 & 347 & 3.28 & 88.3 & 1116 & 1.43 & 1.68
    \\
   &  17.1 & 363.2 & 3.37 & 90 & 1138  & 1.74 & 1.78
    \\
 & 16 & 351 & 3.28 & 89.3 & 1131 & 1.71 & 1.74
    \\
  &18.6 & 354.2 & 3.25 & 91 & 1169 & 1.96 & 1.9
    \\
 & 21.4 & 359.1 & 3.26 & 92 & 1216 & 2.41 & 2.07
    \\
  &  23.1 & 351.3 & 3.18 & 92.2 & 1252  & 2.59 & 2.2
    \\
  &   23.8 & 346.4 & 3.12 & 92.7 & 1200  & 2.62 & 2.04
    \\
   &  13.7 & 348.2 & 3.28 & 88.6 & 1108  & 1.37 & 1.66
    \\
  &   6.2 & 327.1 & 3.32 & 82.3 & 878  & 0.49 & 0.97
    \\
   &9.3 & 334.5 & 3.28 & 85.1 & 1000 & 0.84 & 1.3
    \\
\\
    $160$rev &    17.3 & 383.2 & 3.54 & 90.4 & 1066  & 1.46 & 1.57
    \\
     &14.4 & 367.2 & 3.45 & 88.9 & 975 & 1.25 & 1.29 
    \\
   & 13 & 360.4 & 3.4 & 88.5 & 971 & 1.08 & 1.27
    \\
   &  8.2 & 325.6 & 3.22 & 84.4 & 852  & 0.54 & 0.93
    \\
 & 14 & 346.9 & 3.27 & 88.6 & 962 & 1.18 & 1.25
    \\
  &16.6 & 352.6 & 3.27 & 90 & 1035 & 1.47 & 1.47
    \\
 & 22.9 & 352.5 & 3.19 & 92.3 & 1139 & 2.07 & 1.83
    \\
  &  25.4 & 347.8 & 3.14 & 92.5 & 1113  & 2.25 & 1.75
    \\
  &   23.2 & 346.8 & 3.15 & 91.9 & 1064  & 2.2 & 1.59
    \\
   &  9.5 & 342.4 & 3.35 & 85.3 & 872  & 0.67 & 0.99
    \\
  &   6 & 343.7 & 3.48 & 82.4 & 756  & 0.39 & 0.72
    \\
   & 16.5 & 343.6 & 3.2 & 89.6 & 1039 & 1.4 & 1.48
    \\
\\
    $640$rev & 24.4 & 382.5 & 3.46 & 92.3 & 925  & 1.55 & 1.2
    \\
     & 22.2 & 371.8 & 3.39 & 91.6 & 897 & 1.29 &  1.12
    \\
   & 18.9 & 353.8 & 3.26 & 90.6 & 867 & 1.11 & 1.04
    \\
   &  9.1 & 350 & 3.41 & 85.7 & 682  & 0.42 & 0.61
    \\
 & 7 & 339.9 & 3.38 & 84 & 663 & 0.26 & 0.56
    \\
  &12.1 & 350.5 & 3.35 & 87.3 & 761 & 0.6 & 0.77
    \\
 & 16.8 & 349.8 & 3.24 & 90.1 & 853 & 0.96 & 1
    \\
  &  13.4 & 336.6 & 3.18 & 88.4 & 799  & 0.69 & 0.86 
    \\
  &   17.2 & 332.7 & 3.08 & 90.2 & 865 & 0.97 & 1.03
    \\
   &  22.1 & 355.3 & 3.23 & 91.8 & 892  & 1.28 & 1.11
    \\
  &  25.9 & 356.7 & 3.22 & 92.5 & 894  & 1.61 & 1.13
    \\
   & 8 & 354.9 & 3.49 & 84.9 & 712 & 0.34 & 0.66 
    \\
\hline
    \end{tabular}
    }
    \label{Table:v-blender}
\end{table}

\vspace{-2mm}
\subsubsection{Density}

\hspace{5 mm}All the tablets (made in batch or continuous) were weighed with a precision balance ($\pm0.001$g, Adventurer Ohaus). Their  thickness was carefully measured by a digital caliper ($\pm0.01$mm, Absolute digimatic Caliper). From these measurements, the relative density of the tablets was calculated

\begin{equation}
\bar{\rho} = \frac{\rho_{b}}{\rho_{t}},
\label{eq:RD}
\end{equation}
where $\rho_{b}$ is the bulk density of the tablet and $\rho_{t}$ is the true density of the blend. The thickness, weight, and relative density of all the batch tablets are listed in Tables \ref{Table:labram} and \ref{Table:v-blender}. For continuous tablets, the mean values and standard deviations of all the measured parameters are provided in Table \ref{Table:continuous}. It can be seen that some of the continuous tablets have weights significantly different from the targeted value ($350$mg) which resulted in deviations in their relative density. Detailed information about each individual continuous tablet is provided in the supplementary file.

\begin{table}[htb]
    \centering
\caption{Mean values and standard deviations of weight (W), thickness ($t$), relative density ($\bar{\rho}$), speed of sound (SOS), tensile strength ($\sigma_t$), and Young's modulus (E) of the continuously manufactured tablets compacted at various nominal compaction forces ($\text{F}_n$).}
    \scriptsize\setlength{\tabcolsep}{8pt}{
    \begin{tabular}{c   ccccccc} 
      \hline
         $\text{F}_n$ (kN) & W (mg) & $t$ (mm) & $\bar{\rho}$ (\%) & SOS (m/s) & $\sigma_t$ (MPa) & E (GPa)
    \\
\hline
     8 & 343.9$\pm8$ & 3.35$\pm0.03$ & 85.6$\pm1.3$ & 1147$\pm50$  & 1$\pm0.16$ & 1.72$\pm0.17$ 
    \\

     12 & 350.5$\pm14$ & 3.31$\pm0.07$ & 88.4$\pm1.9$ & 1222$\pm82$ & 1.34$\pm0.44$ & 2.03$\pm0.34$ 
    \\

     16 & 347.8$\pm11.6$ & 3.21$\pm0.07$ & 90.5$\pm1.1$ & 1330$\pm39$ & 1.83$\pm0.28$ & 2.45$\pm0.17$
    \\

     20 & 345.8$\pm8.3$ & 3.13$\pm0.04$ & 92.4$\pm1$ & 1378$\pm27$  & 2.43$\pm0.31$ & 2.68$\pm0.13$
    \\

   24 & 344.9$\pm10.1$ & 3.11$\pm0.06$ & 92.7$\pm0.8$ & 1389$\pm28$ &  2.5$\pm0.31$ & 2.73$\pm0.13$
    \\

  28 & 342.9$\pm5.4$ & 3.05$\pm0.03$ & 94$\pm0.6$ & 1407$\pm26$ &  3.12$\pm0.19$ & 2.84$\pm0.12$
    \\
\hline
    \end{tabular}
    }
    \label{Table:continuous}
\end{table}

\subsubsection{Acoustic Measurements}

\hspace{5 mm}The experimental setup for the US measurements consisted of a pulser/receiver unit (Panametrics, $5077$PR), a pair of protected-face longitudinal wave contact transducers (Panametrics, V$606$-RB) with a central frequency and diameter of $2.25$MHz and $13$mm, respectively, a digitizing oscilloscope (Tektronix TDS$3052$), and a computer controlling the data acquisition. 
During experiments, the pulser/receiver unit was set to a pulse repetition frequency (PRF) of $500$Hz, a pulser voltage of $100$V, an amplification gain of +$10$, a central frequency of $2$-$2.25$MHz, and the entire frequency spectrum of the transducers was allowed to pass through the tablets.

We note that in this study, the US measurements of the continuous tablets were conducted off-line. However, the micro-second TOF of the US signal shows that in principle the US methodology is sufficiently fast to be placed as an in-line measurement.

From the acquired data, the time of flight (TOF) was obtained using the first peak of the received US signal. The SOS was calculated as follows:

\begin{equation}
\text{SOS} = \frac{t}{\text{TOF}},
\end{equation}
where $t$ is the tablet thickness. The US transmission measurement system was tested using steel and aluminum samples with four different thicknesses for each case. A delay time of $1.1$$\mu$s was measured for the setup independent of the material used. Subsequently, this delay time was subtracted from the measured TOF in this study. The Young's modulus of each tablet was then calculated from the SOS assuming the material is isotropic:
\begin{equation}
\text{E} = \text{SOS}^2\rho_{b}.
\label{eq:SOS}
\end{equation}

\subsubsection{Tensile Strength Measurements}

\hspace{5 mm}The diametrical compression test was performed using a standard mechanical hardness tester (Dr. Schleuniger, Pharmatron, model 6D). The tensile strength of tablets, $\sigma_{t}$, is given by \citep{fell1970determination}:

\begin{equation}
\sigma_{t} = \frac{2\text{F}}{\pi \text{D}t},
\label{eq:hertz}
\end{equation}
where F is the breaking force and D is the diameter of the tablet, which is assumed to be constant and equal to $10$mm as radial relaxation was minimal.

\section{A strategy for tensile strength prediction of tablets}

\hspace{5 mm}The mechanical properties of a pharmaceutical tablet not only depend on the material but also on the porosity of the compact. Finding these properties at zero porosity would make it possible to compare blends and predict their maximum strength in compact form. In this study, the fundamental assumption is the existence of a correlation between the Young's modulus and tensile strength at zero porosity of the powder, $\text{E}_{0}$ and $\sigma_{0}$, respectively.
US testing enables us to indirectly measure $\text{E}_{0}$ and, analogously diametrical compression testing allows us to indirectly measure $\sigma_{0}$. These two parameters serve as mechanical characteristics of tablets, which are a function of material properties and processing history. Once we find the relationship between $\text{E}_{0}$ and $\sigma_{0}$, we can predict hardness, using US measurements alone.

In the following sections, the existing theoretical/semi-empirical models to determine $\text{E}_{0}$ and $\sigma_{0}$ are summarized.

\subsection{Young's modulus-porosity correlation}

\hspace{5 mm}The relationship between E and porosity, $\phi$,  is well established in the literature, and both empirical correlations as well as analytical results have been reported. The porosity is defined as $\phi=1-\bar{\rho}$, and \cite{phani1987young} derived a semi-empirical relation with the Young's modulus:

\begin{equation}
\frac{\text{E}}{\text{E}_{0}} = (1-a\phi)^n =f(\phi),
\label{eq:Phani}
\end{equation}
where $a$ and $n$ are material constants, providing information about the packing geometry and pore structure of the material. The constant $a$ is equal to $1/\phi_{c, E}$, where $\phi_{c, E}$ is defined as the porosity of which E vanishes. The minimum value that $a$ can take is 1 corresponding to the maximum value of $\phi_{c, E}=1$. 

\cite{bassam1990young} explored the Young's modulus of fifteen representative tableting excipient powders for different porosities using a four-point beam bending technique and analyzed the data following an empirical function proposed by \cite{spinner1964elastic}:

\begin{equation}
\text{E} = \text{E}_{0}(1-B\phi+C\phi^2),
\label{eq:bassam}
\end{equation}
where $B$ and $C$ are fitting coefficients. Note that Eq. (\ref{eq:bassam}) is equivalent to a second-order Taylor series approximation of Eq. (\ref{eq:Phani}) around zero porosity, with $B=an$ and $C=\frac{a^2}{2} n(n-1)$. The model can be further approximated by a linear regression ($C=0$):

\begin{equation}
\text{E} = \text{E}_{0}(1-m\phi),
\label{eq:Rossi}
\end{equation}
where $m=an$. According to \cite{rossi1968prediction}, $m$ accounts for the stress concentration factor around pores in the material and depends on pore geometry and orientation, which is consistent with the description by \cite{phani1987young}.

\begin{figure}[htb]
\centering
\includegraphics[scale=0.45]{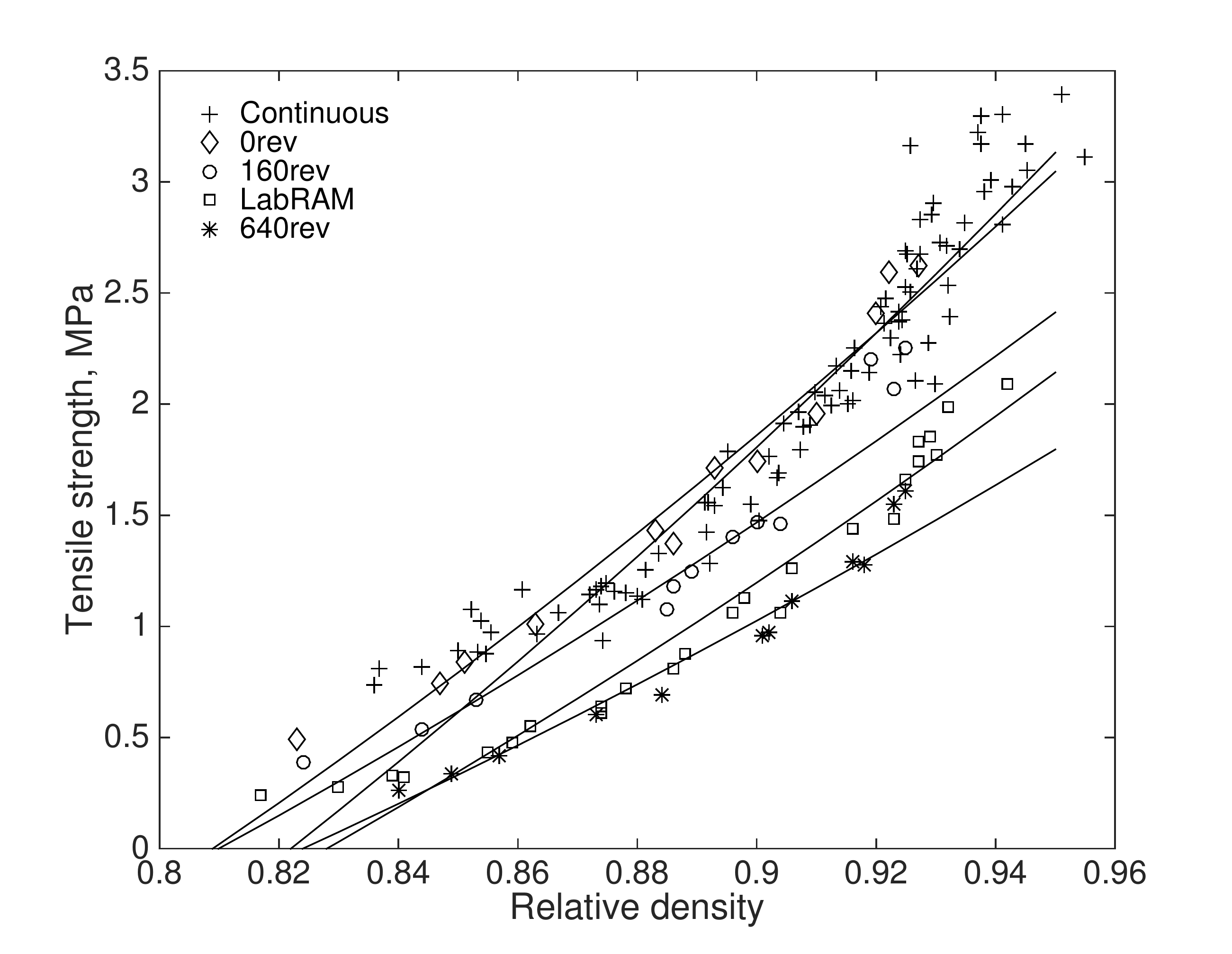}
\caption{The relationship between relative density and tensile strength following the \cite{kuentz2000new} model.}
\label{Fig: RDTSleuenberger}
\end{figure}

\subsection{Tensile strength-porosity correlation}

\hspace{5 mm}We considered a theoretical model based on percolation theory proposed by \cite{kuentz2000new} to relate $\sigma_{t}$ and $\bar{\rho}$:

\begin{equation}
\sigma_{t} = \sigma_{0}[1-(\frac{\phi}{{\phi}_{c, \sigma_{t}}})e^{(\phi_{c, \sigma_{t}}-\phi)}],
\label{eq:leuenberger}
\end{equation}
where $\phi_{c, \sigma_{t}}$ is the porosity at which $\sigma_{t}$ goes to zero.
 
The relationship between $\sigma_{t}$ and $\bar{\rho}$ is generally presented by an exponential form \citep{ryshkewitch1953compression,knudsen1959dependence,tye2005evaluation,razavi2015}. Although this relationship usually provides a good fit to the data, it is empirical and does not provide the limiting values (i.e., $\phi_{c, \sigma_{t}}$ and $\sigma_{0}$) needed to implement the strategy discussed above.

\section{Results and discussion}

\hspace{5 mm}The calculated SOS, $\sigma_{t}$, and E values for continuous and batch tablets are listed in Tables \ref{Table:labram}, \ref{Table:v-blender}, and \ref{Table:continuous}. 

\begin{figure}[htb]
\centering
\includegraphics[scale=0.45]{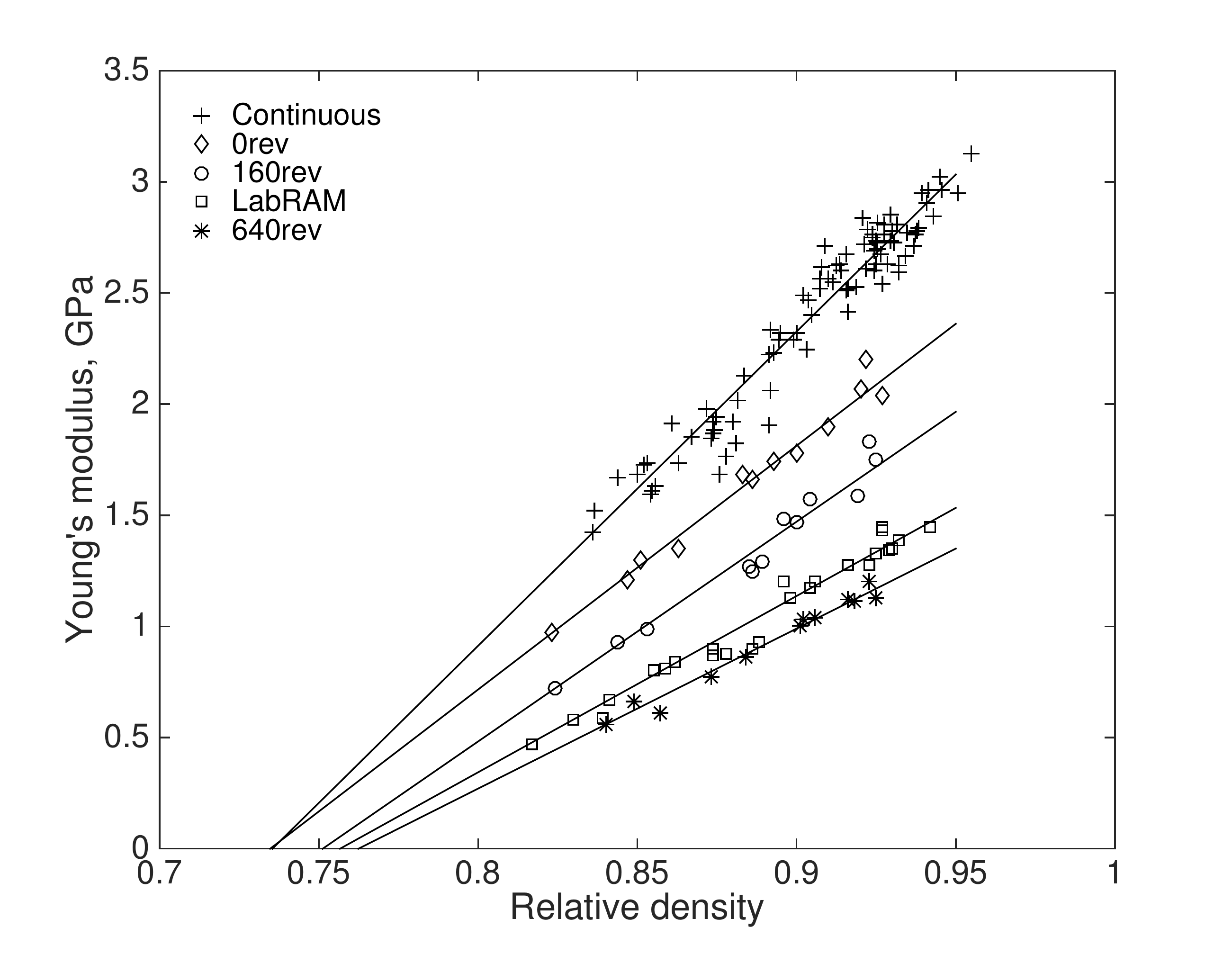}
\caption{Young's modulus as a function of relative density.}
\label{Fig: ERD}
\end{figure}

In Fig. \ref{Fig: RDTSleuenberger}, we present the relationship between $\bar{\rho}$ and $\sigma_{t}$ {\color{red}for} all the continuous and batch tablets. The diametrical compression test results were fitted to Eq. (\ref{eq:leuenberger}) using the non-linear regression method based on the Trust-Region algorithm \citep{MATLAB:2014}. Table \ref{Table:leuenberger} lists the fitted parameters and $r^{2}$ values. We found a reasonable agreement between the model and the experimental results as shown by the $r^{2}$ values. Blends with the same formulation but experiencing different level of shear strain result in different mechanical properties. As expected, the strength of tablets decreased with the level of shear strain.

\begin{table}[htb]
    \centering
\caption{Tensile strength at zero porosity ($\sigma_{0}$) and critical relative density ($\bar{\rho}_{c, \sigma_{t}}$ ) found from Eq. (\ref{eq:leuenberger}) for all the differently produced tablets.}
  \scriptsize\setlength{\tabcolsep}{8pt}{
    \begin{tabular}{c   ccc  ccccccc} 
      \hline
         cases & $\sigma_{0}$ (MPa) & $\bar{\rho}_{c, \sigma_{t}}$ (\%)  & $r^{2}$ 
    \\
\hline
    continuous & 4.60 & 82.2 &0.911
    \\
   0rev &  4.36 & 80.9 & 0.966
    \\
     160rev & 3.46 & 81.0 & 0.921
    \\
     640rev & 2.65 & 82.4 & 0.954
    \\
   labRAM& 3.19 & 82.8 & 0.944
    \\
\hline
    \end{tabular}
    }
    \label{Table:leuenberger}
\end{table}

\begin{table}[htb]
    \centering
\caption{Young's modulus at zero porosity ($\text{E}_{0}$) and other fitting coefficients according to Eqs. (\ref{eq:Phani}), (\ref{eq:bassam}), and (\ref{eq:Rossi}) for all the group of tablets.}
    \scriptsize\setlength{\tabcolsep}{2pt}{
    \begin{tabular}{cc   ccc  ccccccccccccc} 
      \hline
& &&Phani & & &&&&Spinner  & &&&&Rossi
\\
         cases & $\text{E}_{0}$ (GPa)&$a$ & $n$ & $r^{2}$& $\bar{\rho}_{c, E}$ (\%)&&$\text{E}_{0}$ (GPa) & $B$ & $C$ & $r^{2}$&$\bar{\rho}_{c, E}$  (\%)& & $\text{E}_{0}$ (GPa)  & $m$& $r^{2}$&$\bar{\rho}_{c, E}$ (\%)
    \\
\hline
    continuous & 3.73& 3.81 & 0.98 & 0.935 & 73.7& &3.71 & 3.61 &-1.06 &0.935 &74.2&& 3.74&3.78&0.935&73.5
    \\
   0rev & 3.04&3.44&1.22&0.978& 70.9&& 3.06 & 4.28& 2.40&0.978&72.4&&2.91&3.77&0.977&73.5
    \\
     160rev &2.68&3.37&1.46&0.971& 70.3&&2.70 &5.08&5.22&0.971&72.6&&2.46&4.02&0.965&75.1
    \\
     640rev &1.96&2.93&1.98&0.978&65.9&&1.95 & 5.77&8.18&0.978&69.3 &&1.71&4.21&0.978&76.2
    \\
   labRAM& 1.90 & 4.25 & 0.93 & 0.97&74.0&&1.90 & 3.91&-1.13&0.967 &0.76&&1.93&4.11&0.967&75.7
    \\
\hline
    \end{tabular}
    }
    \label{Table:E}
\end{table}

\begin{figure}[htb]
\centering
\includegraphics[scale=0.45]{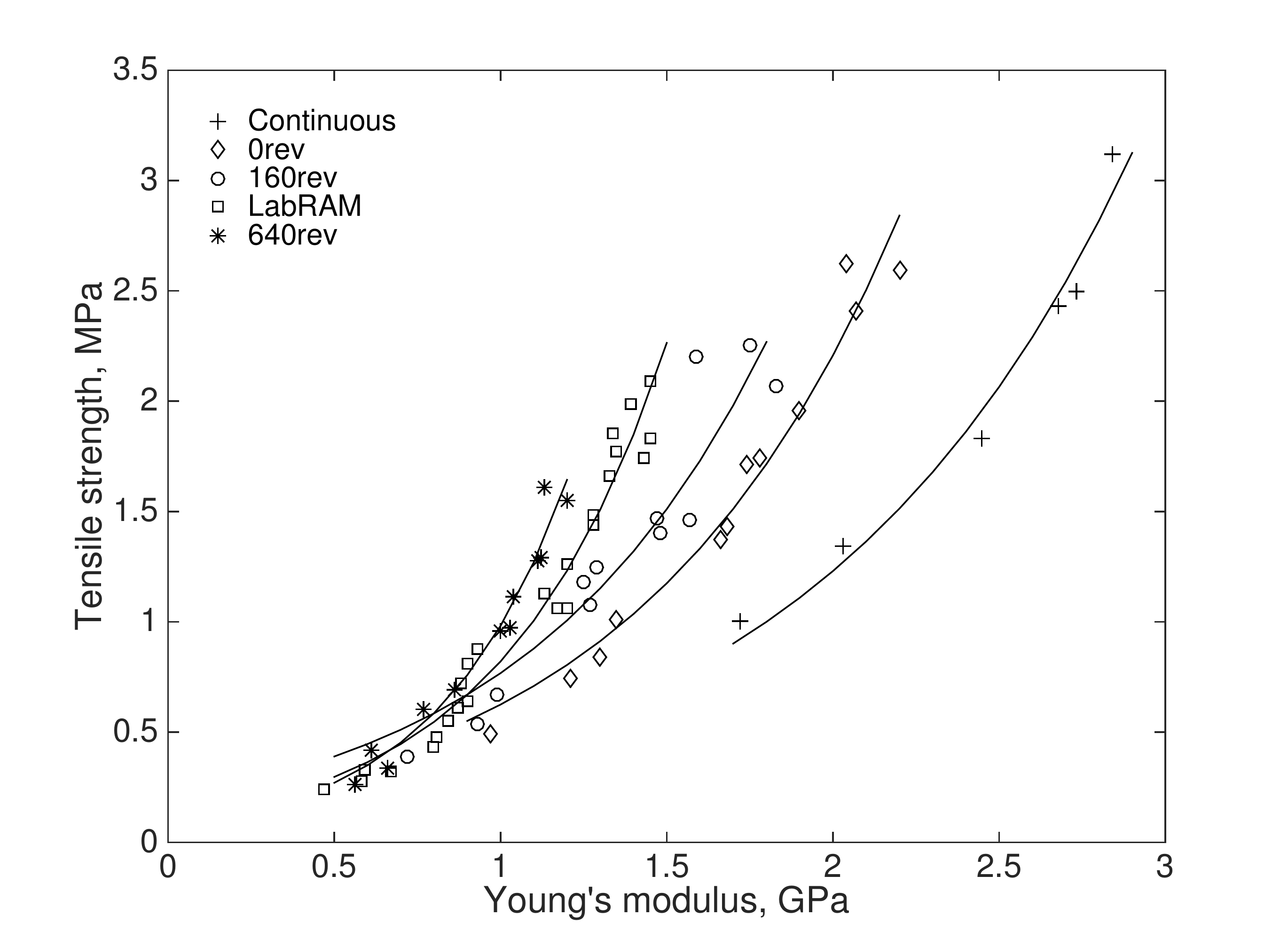}
\caption{Tensile strength as a function of Young's modulus. Mean values are shown for the continuous tablets.}
\label{Fig: ETS}
\end{figure}

\begin{figure}[htb]
\centering
\includegraphics[scale=0.45]{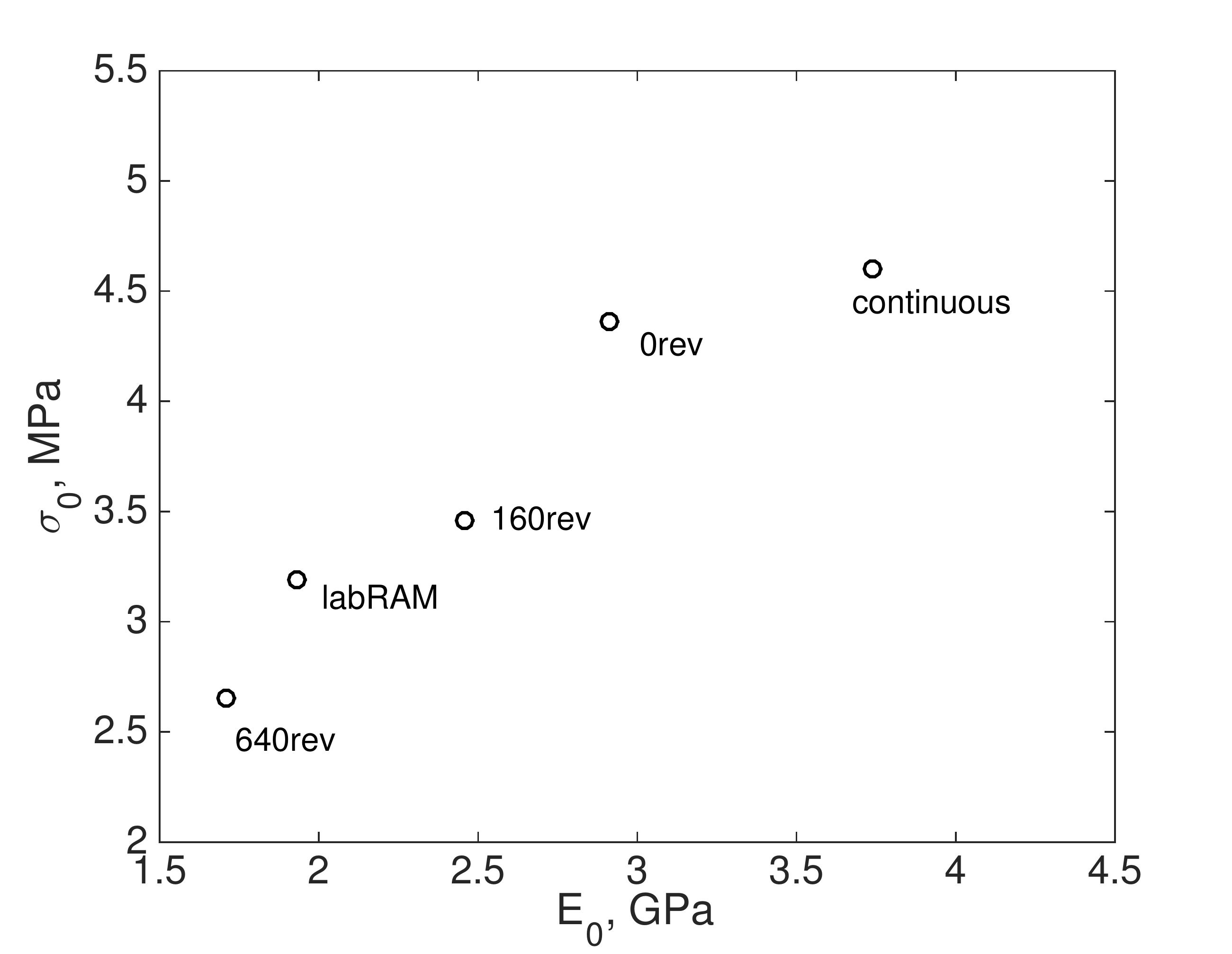}
\caption{A one-to-one relationship between $\text{E}_{0}$ and $\sigma_{0}$ for tablets with the same formulation but different processing history. $\text{E}_{0}$ values are derived from Eq. (\ref{eq:Rossi}). Note that $0$rev, $160$rev, and $640$rev tablets have experienced a 2-minute mixing with MgSt in the V-blender.}
\label{Fig: E0sigma0}
\end{figure}

Fig. \ref{Fig: ERD} shows a linear relationship between $\bar{\rho}$ and E in agreement with  Eq. (\ref{eq:Rossi}). We also fitted the experimental data to Eqs. (\ref{eq:Phani}) and (\ref{eq:bassam}). Table \ref{Table:E} lists the fitted coefficients and $r^{2}$ values of all the tablets. The $\text{E}_{0}$ values are almost the same for all the three models. In addition, we are considering a relatively small range of relative densities and thus, it is acceptable to use a linear regression to estimate $\text{E}_{0}$. It bears emphasis that these results might change for a wider range of relative densities but the methodology would still hold.

It is worth to mention that $\phi_{c, \sigma_{t}}$ and $\phi_{c, E}$ describe two different phenomena, and can be different for the same material and processing conditions. For the material studied here, at a certain relative density we can start to see tensile stiffness in the blend without forming a bond between the particles.

Figs. \ref{Fig: RDTSleuenberger} and \ref{Fig: ERD} allow us to estimate the effective shear strain level corresponding to labRAM and continuous tablets. The selected parameters in labRAM mixer make the hardness of labRAM tablets fall between the $160$rev and $640$rev tablets, indicating that labRAM is a fast and efficient mixing technique \citep{osorio2015evaluation}. On the other hand, the continuous manufacturing process resulted in a low effective shear strain, even below the $0$rev tablets.

It is clear that the total shear strain had an effect on E values. As shear strain increased the SOS values decreased and the tablet{\color{red}s} became both softer and mechanically weaker. The decrease in $\text{E}_{0}$ values with shear strain (Table \ref{Table:E}) is also an indication of weaker bonding between particles.

Based on Figs. \ref{Fig: RDTSleuenberger}  and \ref{Fig: ERD}, US testing seems to be more sensitive than the hardness tester in differentiating tablets. Comparing the continuous and 0rev tablets, for the same $\bar{\rho}$, $\sigma_{t}$ values are very similar. However, their E values are noticeably different.

Fig. \ref{Fig: ETS} shows that we can associate different $\sigma_{t}$ values for one value of E. However, $\bar{\rho}$ values are different for those tablets with the same E and different $\sigma_{t}$. Therefore, knowing both $\bar{\rho}$ and E it is possible to predict tablet hardness. 

Following our proposed strategy, we examine the correlation between the tensile strength and Young's modulus at zero porosity. Fig. \ref{Fig: E0sigma0} shows a one-to-one relationship between $\text{E}_{0}$ and $\sigma_{0}$. Once this relationship is found for a certain blend formulation, we can use US testing only to predict the tensile strength of tablets adopting Eq. (\ref{eq:leuenberger}). This is possible when we know the relative density of tablets prior to US testing.

\section{Conclusion}

\hspace{5 mm}We successfully used ultrasound (US) measurement system on tablets. The effect of two processing parameters, compaction force and level of shear strain were examined. US speed of sound was found to be sensitive to the relative density and the level of shear strain. US testing could detect even small differences between tablets that a hardness tester failed to do so. This is an added advantage of using this technique to monitor mechanical properties, because a slight change could have a significant consequence on dissolution.

A strategy for hardness prediction is proposed that uses the existing models for Young's modulus and tensile strength of porous materials. A clear correlation between Young's modulus and tensile strength at zero porosity is presented. Thus, US testing is a good candidate to be placed on/at-line to measure the mechanical integrity of tablets non-destructively. These results provide information about the behavior of processing parameters on the performance of tablets and the ability to engineer product properties.

\section*{Acknowledgements}

\hspace{5 mm}This material is based upon work supported by the National Science Foundation under Grant number IIP-1237873. The authors wish to thank Golshid Keyvan and Pallavi Pawar for assistance. The authors also gratefully acknowledge the support received from the NSF ERC grant number EEC-0540855.

\bibliographystyle{apalike}
\bibliography{SW}

\end{document}